\documentclass[11pt]{article}
\usepackage{graphicx}
\usepackage{dcolumn}
\usepackage{bm}

\usepackage{fancyhdr}
\usepackage{amsmath} 
\usepackage{graphicx}
\usepackage{epsfig}
\usepackage{tabularx}
\usepackage{amsthm}
\usepackage{amsfonts}
\usepackage{amssymb}
\usepackage{endnotes}
\usepackage{colortbl}
\usepackage{makeidx}
\usepackage{bbold}
\usepackage{bbm}
\usepackage[samesize]{cancel}
\usepackage{mathtools}
\usepackage{color}
\usepackage{enumerate}
\usepackage{relsize}
\usepackage{enumitem} 
\usepackage[top=2cm, bottom=2cm, left=2.cm , right=2.cm]{geometry}
\newcommand{\E}{\mathrm{E}}
\newcommand{\Var}{\mathrm{Var}} 
\newcommand{\Cov}{\mathrm{Cov}}

\begin{document}

\title{Constraint Ornstein-Uhlenbeck bridges}

\author{Alain Mazzolo$^{1}$}


\date{$^{1}$Den-Service d'\'etudes des r\'eacteurs et de math\'ematiques appliqu\'ees (SERMA), CEA, Universit\'e Paris-Saclay, F-91191 Gif-sur-Yvette, France}

\maketitle

\begin{abstract}
In this paper, we study the Ornstein-Uhlenbeck bridge process (i.e. the Ornstein-Uhlenbeck process conditioned to start and end at fixed  points) constraints to have a fixed area under its path. We present both anticipative (in this case, we need the knowledge of the future of the path) and non-anticipative versions of the stochastic process. We obtain the anticipative description thanks to the theory of generalized Gaussian bridges while the non-anticipative representation comes from the theory of stochastic control. For this last representation, a stochastic differential equation is derived  which leads to an effective Langevin equation. Finally, we extend our theoretical findings to linear bridge processes.
\end{abstract}



\section{Introduction}

\label{sec_intro}
The Ornstein-Uhlenbeck process is a diffusion process that was introduced, in a physical context, as a model of the velocity of a particle undergoing a Brownian motion~\cite{ref_book_Mahnke}. Since the original work of Ornstein and Uhlenbeck (1930) the model and its refinements have been used in numerous applications areas. For instance, in biology the Ornstein-Uhlenbeck process models the neuronal activity~\cite{refRicciardi}, in mathematical finance it is used to study stock price distributions~\cite{refStein} and the reflected Ornstein-Uhlenbeck process plays a key role in the context of queuing models with reneging~\cite{refWard} just to name a few. As for Brownian motion, the Ornstein-Uhlenbeck process may be constrained. In particular, with an appropriate drift term the Ornstein-Uhlenbeck process may end at a fixed point $x_f$ at a desired time $T$. Such a constrained Ornstein-Uhlenbeck process is called a generalized Ornstein-Uhlenbeck bridge, or simply a Ornstein-Uhlenbeck bridge when the final point is equal to zero. Not surprisingly, such bridges have also applications in the fields mentioned above. Thus, in biology, the Ornstein-Uhlenbeck bridge process models animal movement in continuous time~\cite{refNiu} and in financial mathematics, the process models the arbitrage process~\cite{refHilliard}.\\

Generalized Ornstein-Uhlenbeck bridges are particular examples of stochastic processes conditioned on their final points. However, more generally, there are several kind of constraints, depending on the physical process studied. For instance, in reactor physics, when the reactor operates at critically the branching process that governs the behavior of the neutron population is subject to two constraints: a constant population (a discrete constraint) and a neutron flux as flat as possible (a constraint related to the spatial extent of the neutron density)~\cite{refdeMulatier}. Another important example, which is still an open problem, comes from computer science where one needs to generate a Brownian bridge conditioned to stay positive with the additional condition of a fixed area under its curve~\cite{refMajumdarCurrSci}. Stochastic processes with a fixed area have also applications in the context of fluctuating interfaces~\cite{refMajumdarComtet,refMajumdarDasgupta}. For others applications of constrained stochastic processes, including the Ornstein-Uhlenbeck bridge, see the revue~\cite{refMajumdarOrland} and the recent article~\cite{ref_LeDoussal} for the (intriguing) non-intersecting Ornstein-Uhlenbeck bridges.\\

However, as for the Brownian case, conditioning a complex object like the Ornstein-Uhlenbeck process is not a harmless task. It is usually achieved thank to Doob's {\it{h}}-transform~\cite{refDoob}. A nice presentation of this method is given in chapter 15 of the book of Karlin and Taylor~\cite{ref_book_Karlin}; it is also outlined, from a physicist point of view, in two recent articles~\cite{refMajumdarOrland,refOrland}. The main ingredient of Doob's method requires the calculation of the probability that from the state value $x$ at time $t$, the sample path of the process satisfies the desired constraint at time $T$. When this quantity is known, Doob's technique has been successfully applied to various kind of conditioned processes~\cite{refMajumdarOrland,ref_book_Karlin,refOrland,refSzavits,refBaudoin,refChetrite}. Unfortunately, a closed-form of this  probability is rarely known, and obtaining its analytic expression often involves special methods. For example, in order to calculate this probability for a Brownian motion ($W_t$) with a fixed area under its curve, we need to consider two processes $A_t=\int_0^t W_s \, ds$ and $W_t$ simultaneously~\cite{refMajumdarComtet,refMajumdarDasgupta}. This is an example of complications that occur when we impose global constraints on the stochastic paths. Nevertheless, due to its Gaussian nature, the distribution of the area under a Brownian bridge is a well-known result, and its extension to various constrained Brownian motions still attracts the attention of the statistical physics community, see~\cite{refSchehr} and references therein. However, standard Brownian motions are often too crude and one has to resort to more elaborated models (Langevin process, L\'evy process, fractional Brownian motion, fractional Ornstein-Uhlenbeck process, etc.) to model real life systems. In reference~\cite{refSchehr}, authors studied the distribution of the area (and its average shape)  under a L\'evy bridge. In the present paper, we follow another generalization by imposing the constraint on the area to an Ornstein-Uhlenbeck bridge. The advantage of the Ornstein-Uhlenbeck bridge is twofold: it is manageable analytically and the Brownian limit can easily be recovered by letting a parameter shrink to zero. At rather little cost, we also consider linear bridge processes. In doing so, we will show that new methods from the applied mathematics literature can be applied successfully to the statistical physics of random processes, thus offering alternative methods for imposing constraints on stochastic processes.\\
 
To overcome the inherent difficulty of Doob's {\it{h}}-transform, several techniques for including global constraints have emerged. In this article, in order to study the problem of a generalized  Ornstein-Uhlenbeck bridge with the additional condition of a fixed area, we will use a couple of recent methods that do not involve Doob's technique. The first one, obtained in the context of generalized Gaussian bridges by Sottinen and Yazigi~\cite{refSottinen} will lead us to an anticipative representation of the conditioned Ornstein-Uhlenbeck bridge process. The second one, due to Chen and Georgiou~\cite{refChen} and based on theory of stochastic control, will lead us to a non-anticipative representation of the stochastic process as well as an effective Langevin equation. \\

\begin{figure}[h]
\centering
\includegraphics[width=4.5in,height=4.in]{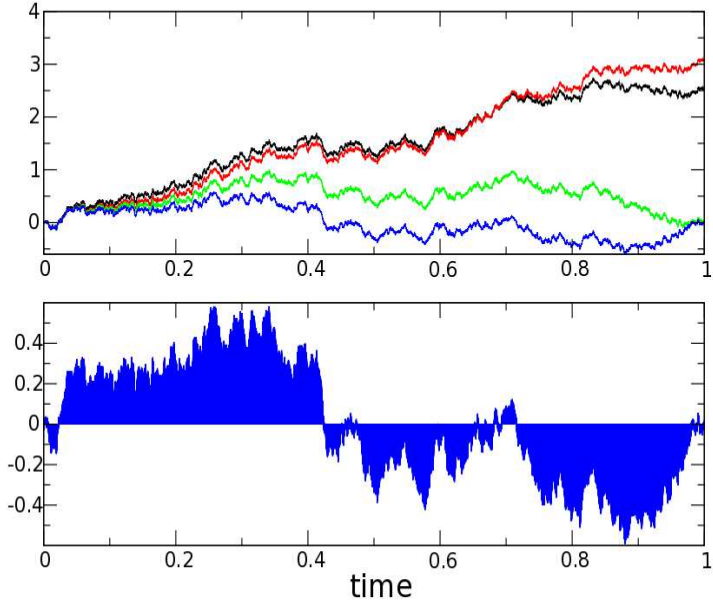}
\setlength{\abovecaptionskip}{15pt} 
\caption{Examples of realizations over the interval $[0,1]$ with the same noise history. Black line: Brownian path. Red line: Ornstein-Uhlenbeck process. Green line: Ornstein-Uhlenbeck bridge. Blue line: Ornstein-Uhlenbeck bridge with a fixed area $A=0$ (zero area Ornstein-Uhlenbeck bridge).}
\label{fig1}
\end{figure}

The paper is organized as follows: in section~\ref{sec2}, after briefly reviewing the Ornstein-Uhlenbeck bridge, we apply the formalism of Sottinen and Yazigi to the generalized Ornstein-Uhlenbeck bridge having a fixed area under its curve to get an anticipative representation of the process. Next, in section~\ref{sec3} a non-anticipative representation of the conditioned process is derived thanks to the theory of stochastic control. In this section,  we also extend our theoretical results to linear bridge processes. A conclusion follows in section~\ref{sec4}. Figure~\ref{fig1} shows some examples of realizations of the various processes studied in this article. \\

\section{Anticipative representation of the constraint Ornstein-Uhlenbeck bridge process}
\label{sec2}
\subsection{Ornstein-Uhlenbeck bridge}
In this section, we apply the formalism of Sottinen and Yazigi~\cite{refSottinen}  to the generalized Ornstein-Uhlenbeck bridge having a fixed area under its curve. This will lead us to the so-called anticipative representation of the process. Before doing so, it may be useful to illustrate the anticipative and non-anticipative representations with the example of the Ornstein-Uhlenbeck bridge. To this end, let us  consider an Ornstein-Uhlenbeck process $X_t$ over the time interval $[0,T]$ described by the stochastic differential equation (SDE)~\cite{ref_book_Mahnke}, 

\begin{equation}
\label{OU-process-SDE}
	dX_t = q X_t dt + \sigma dW_t ,
\end{equation}
where the friction coefficient $q \neq 0$, the diffusion coefficient $\sigma > 0$, and $W_t$ denotes the standard Brownian motion (Wiener process). Assuming that the process begins at $X_0=0$, the solution of the SDE is,
\begin{equation}
\label{OU-solution}
	X_t = \sigma \int_0^t e^{q(t-s)} dW_s,
\end{equation}
a Gaussian process with mean function $\E \left[ X_t \right] =0$ and covariance function $\Cov \left[ X_t X_s \right] = \frac{\sigma^2}{q} e^{q t} \sinh(q s)$ for $0 \leq s \leq t$. From the SDE of the Ornstein-Uhlenbeck process, one can derive the SDE of the generalized Ornstein-Uhlenbeck bridge $\widehat{X}_t $, namely the Ornstein-Uhlenbeck process pinned down at $x_f$ at time $T$. This can be achieved, for instance, thanks to Doob's {\it{h}}-transform~\cite{refMajumdarOrland,refBaudoin} 

\begin{equation}
\label{non-anticipative-OU-bridge}
  \left\{
      \begin{aligned}
        d\widehat{X}_t  &= q \left[- \coth\left[q(T-t)\right] \widehat{X}_t + \frac{x_f}{\sinh\left[q(T-t)\right]} \right] dt + \sigma dW_t \qquad  0 \leq t < T \\
        \widehat{X}_T  &= x_f ,  \\
      \end{aligned}
    \right.
\end{equation}
\noindent whose solution is given by~\cite{refBarczy1}

\begin{equation}
\label{non-anticipative-OU-bridge_sol}
  \widehat{X}_t = \left\{
      \begin{aligned}
        & x_f \frac{\sinh(q t)}{\sinh(q T)} + \int_0^t  \frac{\sinh\left[q(T-t)\right]}{\sinh\left[q(T-s)\right]} \, dW_s \qquad  && \mathrm{if~} 0 \leq t < T \\
        & x_f  \qquad   && \mathrm{if~} t = T.  
      \end{aligned}
    \right.
\end{equation}

\noindent The mathematical literature~\cite{refBarczy1,refBarczy2} gives also an anticipative representation, $\widehat{X}_t ^a$,  of the generalized Ornstein-Uhlenbeck bridge:

\begin{equation}
\label{anticipative-OU-bridge}
   \widehat{X}_t ^a  =  x_f \frac{\sinh(q t)}{\sinh(q T)}  + \left(X_t - \frac{\sinh(q t)}{\sinh(q T)} X_T \right), \qquad  0 \leq t \leq T ,
\end{equation}

\noindent where $X_t$ is a standard Ornstein-Uhlenbeck process. For the details of the various versions of the Ornstein-Uhlenbeck bridge see~\cite{refBarczy2}. The two processes $\widehat{X}_t $ and  $\widehat{X}_t ^a$ have the same law since they have the same expectation and covariance (as it can be easily checked by direct calculation~\cite{refBarczy2}), but to construct the process $\widehat{X}_t ^a$ we need the random variable $X_T$, where the time $T$ is after the time $t$. For this reason, the process $\widehat{X}_t ^a$ is called an anticipative (or non-adapted) representation of the Ornstein-Uhlenbeck bridge. On the contrary, the process $\widehat{X}_t $ is adapted to the filtration $\mathcal{F}_t$ (i.e. "the information at time $t$") generated by $W_t$. This is why the process $\widehat{X}_t $ is called a non-anticipative (or adapted or sometimes canonical~\cite{refSottinen}) representation of the Ornstein-Uhlenbeck bridge. \\

\noindent From now on, we will use the following notations: 
\begin{align*}
	  X_t & : \mathrm{Ornstein-Uhlenbeck~process} \\
	  \widehat{X}_t & : \mathrm{Ornstein-Uhlenbeck~bridge} \\
	  \widehat{\mathbbm{X}}_t & : \mathrm{constrained~Ornstein}-\mathrm{Uhlenbeck~bridge~i.e.~Ornstein-Uhlenbeck~bridge~with~a~fixed~area} \\
	 X_t^a & : \mathrm{the~superscript~"a"~indicates~an~anticipative~version~of~the~process~}X_t
\end{align*}
\noindent The same convention applies to the Brownian motion $W_t$. Thus, $\widehat{W_t}$ designates the Brownian bridge and so on.

\subsection{Constrained Ornstein-Uhlenbeck bridge}
This being specified let us return to the generalized Ornstein-Uhlenbeck bridge having a fixed area under its curve. In the rest of this article, such a process will be referred to as constrained Ornstein-Uhlenbeck bridge. In order to obtain an anticipative version of the process (the equivalent of Eq.~\eqref{anticipative-OU-bridge} for the constrained Ornstein-Uhlenbeck bridge), we employ the recent approach of Sottinen and Yazigi~\cite{refSottinen}, a technique that is particularly suited to include global constraints on stochastic paths. We have already used their theory in a recent article on the constrained Brownian bridge~\cite{refMazzolo} and in this section we pursue our efforts with the constrained Ornstein-Uhlenbeck process. In the following, we use the notation of Sottinen and Yazigi by naming $R[s,t]$ the covariance of the process $X_t$, i.e.,
\begin{equation} 
\label{covariance}
 	R(t,s) := \Cov \left[X_t  \,  X_s \right] = \E \left[ \left( X_t - \E[X_t] \right) \left( X_s - \E[X_s] \right)\right].
\end{equation}

\noindent Sottinen and Yazigi's result is valid for any zero mean Gaussian process and therefore for the Ornstein-Uhlenbeck
 process we are interested in. It states that for such a process, the anticipative representation of the process conditioned on its final value $x_f$ and on the area $A$ under its path, 
\begin{equation}
\label{two_constraints}
  \left\{
      \begin{aligned}
        & \widehat{\mathbbm{X}}_T^a  = x_f \\
        & \int_0^T \widehat{\mathbbm{X}}_t^a  \, dt  = A,  \\
      \end{aligned}
    \right.
\end{equation}
\noindent is a Gaussian process given by~\cite{refSottinen}

\begin{align}
	\widehat{\mathbbm{X}}_t^a  =  X_t & - \frac{R(t,T) \int_0^T \int_0^T R(s,u) \, ds \, du  - \int_0^T R(t,s) \, ds \int_0^T R(T,u) \, du }{R(T,T) \int_0^T \int_0^T R(s,u) \, ds \, du - \int_0^T R(T,s) \, ds \int_0^T R(T,u) \, du  } \left(X_T -x_f \right) \nonumber \\
           & -  \frac{R(T,T) \int_0^T R(s,t) \, ds  - R( t,T) \int_0^T R(T,s) \, ds }{R(T,T) \int_0^T \int_0^T R(s,u) \, ds \, du - \int_0^T R(T,s) \, ds \int_0^T R(T,u) \, du  } \left( \int_0^T (T-t) \, dX_t - A  \right).
\end{align}

\noindent The stochastic integral, thanks to It\={o}'s lemma $d[(T-t)X_t] = - X_t dt + (T-t) dX_t$, can be transformed into a regular one and we immediately get

\begin{align}
\label{constrained-anticipative-bridge}
	\widehat{\mathbbm{X}}_t^a   =  X_t & + \frac{R(t,T) \int_0^T \int_0^T R(s,u) \, ds \, du  - \int_0^T R(t,s) \, ds \int_0^T R(T,u) \, du }{R(T,T) \int_0^T \int_0^T R(s,u) \, ds \, du - \int_0^T R(T,s) \, ds \int_0^T R(T,u) \, du  } \left(x_f - X_T \right) \nonumber \\
           & +  \frac{R(T,T) \int_0^T R(s,t) \, ds  - R( t,T) \int_0^T R(T,s) \, ds  }{R(T,T) \int_0^T \int_0^T R(s,u) \, ds \, du - \int_0^T R(T,s) \, ds \int_0^T R(T,u) \, du  } \left(A  - \int_0^T X_t  \, dt \right).
\end{align}
 
\noindent Equation~\eqref{constrained-anticipative-bridge} is an anticipative representation of a stochastic process pinned down to the value $x_f$ at time $T$ and constrained to have a fixed area $A$ under its curve. From the structure of this equation, it is straightforward to check that the process satisfies both constraints. The first constraint $\widehat{\mathbbm{X}}_T^a =x_f$ is obvious and integrating Eq.~\eqref{constrained-anticipative-bridge} over $[0,T]$ yields,

\begin{align}
\label{anticipative-bridge-integrated}
	\int_0^T  \widehat{\mathbbm{X}}_t^a   \, dt  = \int_0^T  X_t \, dt & + \underbrace{ \frac{\int_0^T R(t,T) \, dt \int_0^T \int_0^T R(s,u) \, ds \, du  - \int_0^T \int_0^T R(t,s) \, ds \, dt \int_0^T R(T,u) \, du }{R(T,T) \int_0^T \int_0^T R(s,u) \, ds \, du - \int_0^T R(T,s) \, ds \int_0^T R(T,u) \, du  } }_{= 0} \left(x_f - X_T \right) \nonumber \\
          &  +  \underbrace{ \frac{R(T,T) \int_0^T \int_0^T R(s,t) \, ds \, dt - \int_0^T R( t,T) \, ds \int_0^T R(T,s) \, ds  }{R(T,T) \int_0^T \int_0^T R(s,u) \, ds \, du - \int_0^T R(T,s) \, ds \int_0^T R(T,u) \, du  }}_{= 1}  \left(A  - \int_0^T X_t  \, dt \right) = A.
\end{align}

\noindent Moreover, since the covariance of the Ornstein-Uhlenbeck process is given by,
\begin{equation}
\label{OU-covariance}
	R(t,s) = \Cov \left[X_t  \,  X_s \right] = \frac{\sigma^2}{2q} e^{q(t+s)} \left(1 - e^{-2q \min(t,s)} \right),
\end{equation}
\noindent all the integrals in Eq.\eqref{constrained-anticipative-bridge} can be straightforwardly derived from the preceding expression. Once this is done, we obtain,

\begin{align}
\label{constrained-OU-bridge-anticipative}
	\widehat{\mathbbm{X}}_t^a   =  X_t & + \frac{(1-e^{-qt})\left(e^{qt}-e^{2qT}+e^{q(T+t)}(qT-1)+e^{qT}(qT+1) \right)}{(e^{qT}-1)(2+qT+e^{qT}(qT-2))} \left(x_f - X_T \right) \nonumber \\
           & +  \frac{2q \sinh \left(\frac{qt}{2} \right) \sinh \left(\frac{q}{2}(T-t) \right) } {qT \cosh \left(\frac{qT}{2} \right)  - 2 \sinh \left(\frac{qT}{2} \right)} \left(A  - \int_0^T X_t  \, dt \right).
\end{align}

\noindent Equation~\eqref{constrained-OU-bridge-anticipative} is an anticipative representation of a constrained Ornstein-Uhlenbeck bridge (fixed area $A$). As we have observed, constraints on the stochastic process appear clearly in an anticipative version and in addition this representation has one other great advantage:  averaging Eq.~\eqref{constrained-OU-bridge-anticipative} over the realizations gives the mean behavior of the process very easily. Indeed, recall that since the Ornstein-Uhlenbeck process starts at $X_0=0$, we have $\E \left[  X_t \right] =0$ and consequently averaging Eq.~\eqref{constrained-OU-bridge-anticipative} yields immediately to,

\begin{equation}
\label{constrained_OU-bridge-anticipative-mean}
	\E  [\widehat{\mathbbm{X}}_t^a ] =  \frac{(1-e^{-qt})\left(e^{qt}-e^{2qT}+e^{q(T+t)}(qT-1)+e^{qT}(qT+1)\right)} {(e^{qT}-1)(2+qT+e^{qT}(qT-2))} x_f + \frac{2q \sinh \left(\frac{qt}{2} \right) \sinh \left(\frac{q}{2}(T-t) \right)} {qT \cosh \left(\frac{qT}{2} \right)  - 2 \sinh \left(\frac{qT}{2} \right)} A .
\end{equation}

\begin{figure}[!h]
\centering
\includegraphics[width=4.5in,height=3.2in]{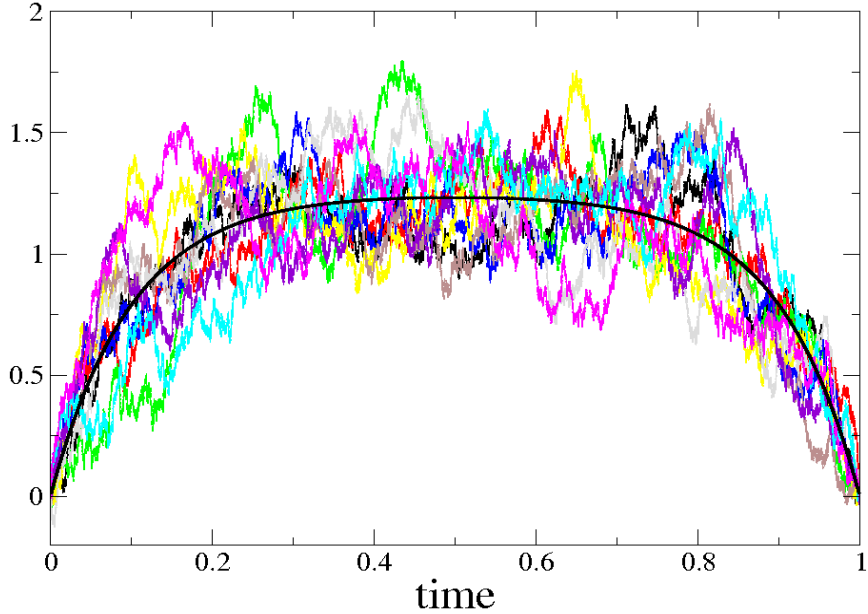}
\setlength{\abovecaptionskip}{15pt} 
\caption{A sample of 10 Ornstein-Uhlenbeck processes with $q= 10$ and $\sigma =1$ conditioned to end at $x_f = 0$ and having a unit area. The time step used in the discretization is $dt=10^{-5}$. All trajectories are statistically independent. The thick black curve is the average profile of the stochastic process as given by Eq.(\ref{constrained_OU-bridge-anticipative-mean}).}
\label{fig2}
\end{figure}

\noindent The mean behavior of the generalized constrained Ornstein-Uhlenbeck bridge is significantly more complicated than that of the Ornstein-Uhlenbeck bridge which reads, after averaging Eq.~\eqref{anticipative-OU-bridge} over the realizations,

\begin{equation}
\label{OU-bridge-anticipative-mean}
	\E [X_t ] = x_f \frac{\sinh(q t)}{\sinh(q T)} .
\end{equation}
\noindent Figure~\ref{fig2} shows a set of 10 realizations of the process ending at $x_f = 0$ at time $T = 1$ with a fixed area of $1$ as well as its mean trajectory. \\

\vspace{1cm}

\noindent To complete the characterization of the Gaussian process, we also have to calculate its variance. To this end, let us first simplify the notation of Eq.~\eqref{constrained-OU-bridge-anticipative} by defining two non-random functions $\alpha(t)$ and $\beta(t)$:

\begin{equation}
\label{definition_functions_a_and_b}
  \left\{
      \begin{aligned}
        & \alpha(t) := \frac{(1-e^{-qt})\left(e^{qt}-e^{2qT}+e^{q(T+t)}(qT-1)+e^{qT}(qT+1) \right)}{(e^{qT}-1)(2+qT+e^{qT}(qT-2))} \\
        & \beta(t) :=  \frac{2q \sinh \left(\frac{qt}{2} \right) \sinh \left(\frac{q}{2}(T-t) \right) } {qT \cosh \left(\frac{qT}{2} \right)  - 2 \sinh \left(\frac{qT}{2} \right)} , \\
      \end{aligned}
    \right.
\end{equation}

\noindent so that Eq.~\eqref{constrained-OU-bridge-anticipative} simplifies to:

\begin{equation}
\label{constrained-OU-bridge-anticipative-simplified}
	\widehat{\mathbbm{X}}_t^a  =  X_t  + \alpha(t) \left(x_f - X_T \right) + \beta(t) \left(A  - \int_0^T X_s  \, ds \right), 
\end{equation}

\noindent and the mean behavior becomes, $\E [\widehat{\mathbbm{X}}_t^a ] =  \alpha(t) x_f + \beta(t) A $. With this notation, the variance writes

\begin{align}
\label{variance-constrained-OU-bridge-anticipative-start}
	\Var [\widehat{\mathbbm{X}}_t^a ] & = \E \left[ \left( X_t  - \alpha(t) X_T  - \beta(t) \int_0^T X_s  \, ds  \right)^2 \right] \nonumber \\
           & = R(t,t) + {\alpha}^2(t) R(T,T) + {\beta}^2(t) \int_0^T \int_0^T  R(u,s) \, du \, ds \nonumber \\
           &  - 2 \alpha(t) R(t,T) - 2 \beta(t) \int_0^T R(t,s) \, ds + 2 \alpha(t) \beta(t)  \int_0^T R(T,s) \, ds \, .
\end{align}

\noindent We have already encountered these kinds of integrals previously. Performing them and replacing $\alpha(t)$ and $\beta(t)$ by their analytical expressions in~Eq.\eqref{variance-constrained-OU-bridge-anticipative-start} leads, after straightforward but lengthy calculations, to

\begin{align}
\label{variance-constrained-OU-bridge-anticipative-final}
	\Var [\widehat{\mathbbm{X}}_t^a ] & = \frac{\sigma ^2 \left(e^{q t}-1\right) e^{-q (3 t+T)} }{2 q \left(e^{q T}-1\right) \left(q T+e^{q T} (q T-2)+2\right)^2} \left(4 (q T-2) e^{3 q (t+T)}-q T (q T-2) e^{2 q (2 t+T)}+q T (q T-2) e^{q (t+4 T)} \right. \nonumber \\
     & +(q T-4) (q T-2) e^{2 q (t+2 T)}+ 4 (q T+2) e^{2 q (t+T)}-q T (q T+2) e^{q (4 t+T)}+q T (q T+2) e^{q (t+3 T)}  \nonumber \\
     & \left. -(q T+2) (q T+4) e^{q (3 t+T)}+(q T (q T+2)-16) e^{q (2 t+3 T)}+(q T (2-q T)+16) e^{q (3 t+2 T)}  \right)  .  
\end{align}

\noindent A rather cumbersome expression compared to that of the standard Ornstein-Uhlenbeck bridge variance~\cite{refBarczy2},
\begin{equation}
\label{variance-OU-bridge}
	\Var [\widehat{X}_t^a]  = \frac{\sigma^2}{q} \frac{\sinh(q t) \sinh(q(T-t))}{\sinh(q T)} .
\end{equation}
Remark that the variance of the constrained Ornstein-Uhlenbeck bridge is independent of the constraints, $x_f$ and $A$ since these constraints act only on the non-random functions $\alpha(t)$ and $\beta(t)$ (Eq.~\eqref{constrained-OU-bridge-anticipative-simplified}). \\

\begin{figure}[!h]
\centering
\includegraphics[width=4.2in,height=4.in]{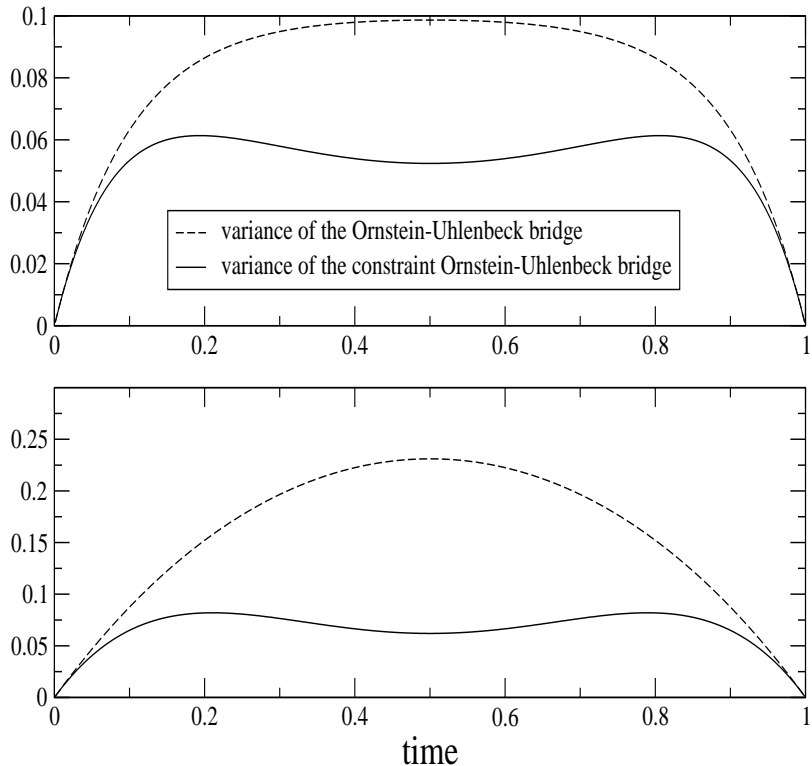}
\setlength{\abovecaptionskip}{15pt} 
\caption{Variance of the Ornstein-Uhlenbeck brigde $\widehat{X}_t^a$ (dash line) and of the constraint Ornstein-Uhlenbeck brigde  $\widehat{\mathbbm{X}}_t^a$ (solid line) for two different Ornstein-Uhlenbeck processes: $q = 1$, and  $q =5 $, from top to bottom. The variance of $\widehat{X}_t^a$ is given by Eq.(\ref{variance-OU-bridge}) and that of $\widehat{\mathbbm{X}}_t^a$ by Eq.(\ref{variance-constrained-OU-bridge-anticipative-final}). Compared with the Ornstein-Uhlenbeck bridge, the additional constraint on the area has the effect of reducing the variance as expect intuitively. Indeed, the more the process is constrained the less it has the possibility to deviate from its average value. Moreover, both variances equal zero at the final time since both processes end at a desired point $x_f$ with probability $1$. Note that the two maxima and the minimum at $1/2$ are not due to the Ornstein-Uhlenbeck process since they also exist in the Brownian limit, when $q \to 0$. Indeed, the variance of the constrained Brownian bridge given by Eq.~\eqref{variance-constrained-Brownian-bridge} has a minimum at $T/2$ and two maxima at $(3-\sqrt{3})T/6$ and $(3+\sqrt{3})T/6$, both at equal distance from the extremities.}
\label{fig3}
\end{figure}

\noindent Figure~\ref{fig3} shows the behavior of the variance of the Ornstein-Uhlenbeck bridge and that of the conditioned Ornstein-Uhlenbeck bridge for two different values of the friction coefficient. Remark that when $t$ approaches $T$, $\Var [\widehat{\mathbbm{X}}_t^a ] \simeq \sigma^2 (T-t) + o(T-t)^2$ and thus $\Var [\widehat{\mathbbm{X}}_T^a ] = 0$ as expected since the process ends at a fixed point $x_f$ at time $T$ with probability 1. Note also that the small $q$ limit gives the variance of the constrained Brownian bridge process. Indeed, from Eq.\eqref{variance-constrained-OU-bridge-anticipative-final} we immediately get, \\

\begin{equation}
\label{variance-constrained-Brownian-bridge}
	\lim_{q \to 0} \Var [\widehat{\mathbbm{X}}_t^a ] =  \Var [\widehat{\mathbbm{W}}_t^a ] = \sigma ^2 \frac{t}{T^3} (T-t) \left(3 t^2 -3 t T +t^2 \right)  .
\end{equation}\\

\noindent Moreover, from  Eq.~\eqref{OU-process-SDE} or  Eq.~\eqref{OU-solution}, we know that when $q \to 0$ and $\sigma =1$, the Ornstein-Uhlenbeck process reduces to a standard Brownian motion ($\lim_{q \to 0} X_t = W_t$). In this limit, we have,

\begin{equation} 
	\lim_{q \to 0} \alpha(t) = \lim_{q \to 0}  \frac{(1-e^{-qt})\left(e^{qt}-e^{2qT}+e^{q(T+t)}(qT-1)+e^{qT}(qT+1) \right)}{(e^{qT}-1)(2+qT+e^{qT}(qT-2))} = \frac{t (3 t - 2 T)}{T^2},
\end{equation}
\noindent and,
\begin{equation}
	\lim_{q \to 0} \beta(t) = \lim_{q \to 0} \frac{2q \sinh \left(\frac{qt}{2} \right) \sinh \left(\frac{q}{2}(T-t) \right) } {qT \cosh \left(\frac{qT}{2} \right)  - 2 \sinh \left(\frac{qT}{2} \right)}= \frac{6 t (T-t)}{T^3}.
\end{equation}

\noindent Replacing these two limits in the Eq.~\eqref{constrained-OU-bridge-anticipative-simplified} leads to,
\begin{equation}
\label{anticipative-Brownian-bridge}
	\widehat{\mathbbm{W}}_t^a  =  W_t + \frac{t (3 t - 2 T)}{T^2} \left(x_f - W_T \right)  + \frac{6 t (T-t)}{T^3} \left(A - \int_0^T W_s  \, ds \right) ,
\end{equation}
\noindent which is an anticipative representation of a generalized Brownian bridge conditioned to have a fixed area. This equation was recently derived in~\cite{refGorgens,refMazzolo}. Thus, as expected, the anticipative version of the generalized constrained Ornstein-Uhlenbeck bridge is in accordance with the corresponding anticipative version of the generalized constrained Brownian bridge. A similar behavior between the Ornstein-Uhlenbeck bridge and the Brownian bridge (thus without other constraint) was obtained in~\cite{refBarczy2}.

\subsection{Constrained Ornstein-Uhlenbeck bridge with drift}
In this paragraph, we consider the Ornstein-Uhlenbeck process with an additional drift term $r(t)$. We will study the influence of this additional term on the behavior of the constrained process. In order to differentiate the original processes without drift and the processes with drift we add a star to the processes with drift, so $X_t^*$ designates an Ornstein-Uhlenbeck process with drift and $\widehat{\mathbbm{X}}_t^{*} $ the constrained Ornstein-Uhlenbeck bridge associated with the drifted Ornstein-Uhlenbeck process. The drifted Ornstein-Uhlenbeck process satisfies the stochastic differential equation,

\begin{equation}
\label{OU-process-SDE-with-drift}
	dX_t^* = (q X_t^* + r(t)) dt + \sigma dW_t ,
\end{equation}

\noindent whose solution is (again we assume that the process starts at zero, $X_0^* = 0$),

\begin{equation}
\label{OU-process-SDE-with-drift-solution}
	X_t^* =  \int_0^t e^{q(t-s)} r(s)\, ds + \sigma \int_0^t e^{q(t-s)} \, dW_s .
\end{equation}

\noindent Since the mean is $\E \left[ X_t^* \right] =  \int_0^t e^{q(t-s)} r(s)\, ds $, $X_t^* - \E \left[ X_t^* \right] = X_t$ where $X_t$ denotes, as usual, the Ornstein-Uhlenbeck process without drift. Thus, $\Cov[X_t^* X_s^*]= \E \left[(X_t^* - \E \left[ X_t^* \right])(X_s^*  - \E \left[ X_s^*  \right])\right]=R(t,s)$ and in Eq.~\eqref{constrained-anticipative-bridge} all integrals involving $R(t,s)$ are left unchanged, as are the functions $\alpha(t)$ and $\beta(t)$ which are a combination of these integrals. So, the anticipative version of the drifted Ornstein-Uhlenbeck process conditioned to end at $x_f$ and conditioned to have a fixed area $A$ is given by,

\begin{equation}
\label{constrained-OU-bridge-with-drift-anticipative-simplified}
	\widehat{\mathbbm{X}}_t^{*a}   =  X_t^{*}  + \alpha(t) \left(x_f - X_T^{*} \right) + \beta(t) \left(A  - \int_0^T X_s^{*}  \, ds \right)
\end{equation}

\noindent or by replacing $X_t^* = X_t + \E \left[ X_t^* \right]$ in the previous expression,

\begin{align}
\label{constrained-OU-bridge-with-drift-anticipative-simplified-2}
	\widehat{\mathbbm{X}}_t^{*a}  & =   X_t  + \alpha(t) \left(x_f - X_T \right) + \beta(t) \left(A  - \int_0^T X_s  \, ds \right)  \nonumber \\
              & + \E \left[X_t^{*} \right] - \alpha(t) \, \E \left[X_T^{*} \right] -\beta(t) \int_0^T \E \left[X_s^{*} \right]\, ds .
\end{align}

\noindent The first line in the r.h.s. of Eq.~\eqref{constrained-OU-bridge-with-drift-anticipative-simplified-2} is the anticipative representation of a generalized Ornstein-Uhlenbeck bridge constrained to have a fixed area $A$ under its path Eq.~\eqref{constrained-OU-bridge-anticipative-simplified}. Thus, the drift term has the effect of adding a deterministic function (given by the second line in the r.h.s. of Eq.~\eqref{constrained-OU-bridge-with-drift-anticipative-simplified-2}).
Since the functions $\alpha(t)$ and $\beta(t)$ are known (Eq.~\eqref{definition_functions_a_and_b}), switching from the case without drift to the case with drift requires only the calculation of $\E \left[X_t^{*} \right]$ and its integral over time. To illustrate this, we consider two examples: i) a constant drift, ii) a drift with a singularity at the final time of the form $r/\sqrt{T-t}$.

\begin{figure}[!h]
\centering
\includegraphics[width=4.5in,height=3.2in]{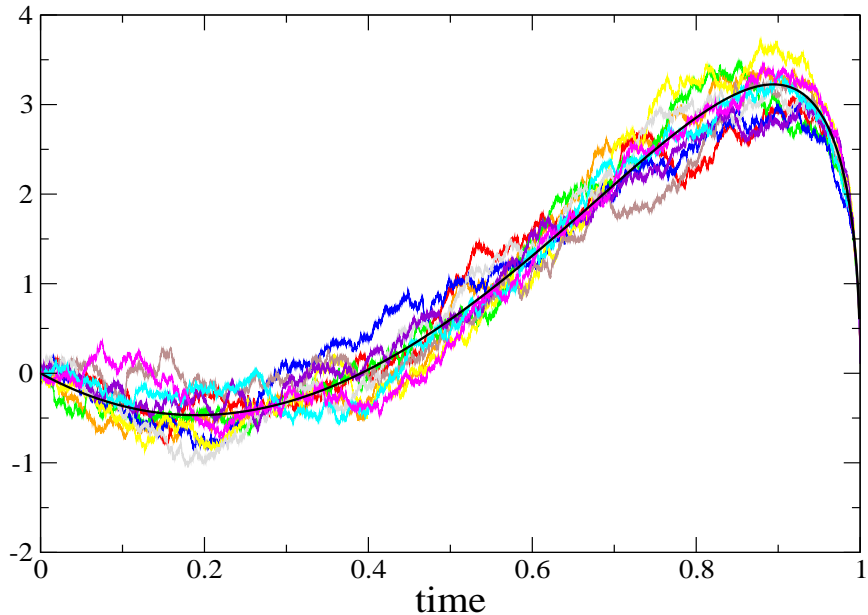}
\setlength{\abovecaptionskip}{15pt} 
\caption{A sample of 10 Ornstein-Uhlenbeck processes with $q= 1$ and $\sigma =1$ with an additional drift term given by $-10/\sqrt{T-t}$. The process is conditioned to end at $x_f = 0$ and conditioned to have a fixed area $A=1$. The time step used in the discretization is $dt=10^{-5}$. All trajectories are statistically independent. The thick black curve is the average profile of the stochastic process as given by Eq.(\ref{constrained-OU-bridge-with-odd-drift-anticipative-mean}).}
\label{fig4}
\end{figure}

\begin{enumerate}[label=\roman*)]
\item By taking a constant drift $r(t)=r \ne 0$, one obtains, $\E \left[X_t^{*} \right] = \frac{r}{q} \left(e^{q t} -1 \right)$ and $\int_0^T \E \left[X_s^{*} \right]\, ds = \frac{r}{q^2} \left(e^{q T} -1 - q T \right) $. With this particular set of values, we have the remarkable relation between $\alpha(t)$, $\beta(t)$ and $\E \left[X_t^{*} \right]$ ($\alpha(t)$ and $\beta(t)$, left unchanged, are given by Eq.~\eqref{definition_functions_a_and_b}):
\begin{equation}
\label{remarkable_relation}
	\E \left[X_t^{*} \right] - \alpha(t) \E \left[X_T^{*} \right] - \beta(t) \int_0^T \E \left[X_s^{*} \right]\, ds = 0 \, .
\end{equation}

\noindent An immediate consequence of this relation is that $\widehat{\mathbbm{X}}_t^{*a} =   X_t  + \alpha(t) \left(x_f - X_T \right) + \beta(t) \left(A  - \int_0^T X_s  \, ds \right)$ which means that the conditioned process with drift has the same law as the conditioned process without drift. In others words, adding a constant drift to the original Ornstein-Uhlenbeck process has no effect on the conditioning.

\item By taking a drift of the form $r(t)=r/\sqrt{T-t}$, one obtains, $\E \left[X_t^{*} \right] = \frac{2 r}{\sqrt{q}} e^{q t} D(\sqrt{q T})$ and $\int_0^T \E \left[X_s^{*} \right]\, ds = \frac{2 r}{q^{3/2}} \left( e^{q T}D(\sqrt{q T}) - \sqrt{q T} \right) $, where $D(x) := e^{-x^2} \int_0^x e^{t^2} \, dt $ is the Dawson function. Replacing these expressions in Eq.~\eqref{constrained-OU-bridge-with-drift-anticipative-simplified-2} leads to,
\begin{align}
\label{constrained-OU-bridge-with-odd-drift-anticipative-simplified}
	\widehat{\mathbbm{X}}_t^{*a} & =   X_t  + \alpha(t) \left(x_f - X_T \right) + \beta(t) \left(A  - \int_0^T X_s  \, ds \right)  \nonumber \\
              & + \frac{2 r}{\sqrt{q}} \left[ e^{q t} D(\sqrt{q T}) - \alpha(t) \, e^{q T} D(\sqrt{q T})  - \frac{\beta(t)}{q} \left( e^{q T}D(\sqrt{q T}) - \sqrt{q T} \right) \right]  ,
\end{align}
\noindent which is the anticipative version of the constrained Ornstein-Uhlenbeck bridge with a deterministic drift of the form $r/\sqrt{T-t}$. Averaging this equation over the realizations gives immediately the mean behavior of the process,  
\begin{equation}
\label{constrained-OU-bridge-with-odd-drift-anticipative-mean}
	\E [\widehat{\mathbbm{X}}_t^{*a} ] = \alpha(t) x_f  + \beta(t) A + \frac{2 r}{\sqrt{q}} \left[ e^{q t} D(\sqrt{q T}) - \alpha(t) \, e^{q T} D(\sqrt{q T})  - \frac{\beta(t)}{q} \left( e^{q T}D(\sqrt{q T}) - \sqrt{q T} \right) \right]  .
\end{equation}
\noindent Figure~\ref{fig4}) shows a set of 10 realizations of the process ending at $T=1$ with a fixed area of $1$ as well as its mean trajectory. Due to the shape of the drift, the behavior of the process differs drastically from that of the driftless process (plotted on Figure~\ref{fig2}). The conditioning affects the Ornstein-Uhlenbeck process substantially at the boundary of the time interval $[0,T]$. In particular, the process is no longer symmetrical with respect to time contrary to the Ornstein-Uhlenbeck bridge.\\
\end{enumerate}

\vspace{0.5cm}

\noindent Despite its advantages, an anticipative version of the conditioned process requires the knowledge of a point $X_T$ in the future and is thus not entirely satisfactory. Having a non-anticipative representation is therefore crucial. To this aim, we first thought of applying Sottinen and Yazigi's formalism as we did recently with the constrained Brownian motion~\cite{refMazzolo}, but since the underlying process is the Ornstein-Uhlenbeck process this approach leads to a stochastic differential equation with a $dX_t$ term instead of a pure Brownian term $dW_t$. As such, obtaining a Langevin equation is not straightforward. Fortunately, very recently, based on the theory of stochastic control, Chen and Georgiou~\cite{refChen} have developed an elegant method that handles the constraints we wish to impose on the Ornstein-Uhlenbeck process. Their method leads to a stochastic differential equation description of the conditioned process, and turns out to be amazingly simple. The rest of the paper is devoted to their technique, first to the constraint Ornstein-Uhlenbeck process and then to the constraint linear process.

\section{Non-anticipative representation of the constraint Ornstein-Uhlenbeck bridge process}
\label{sec3}
\subsection{Conditioning by the stochastic control approach}
In a recent article, based on ideas from optimal stochastic control, Chen and Georgiou~\cite{refChen} have derived a stochastic differential equation for an Ornstein-Uhlenbeck process, and more generally for a linear bridge process conditioned on their positions and velocities at the initial and final times. Let us outline their results in this general framework that encompasses the Brownian and the Ornstein-Uhlenbeck bridges. For this purpose, we introduce the vectorial process $\bm{\xi}_t$ having two components $A_t$ and $X_t$,

\begin{equation}
\label{definition_xi}
	\bm{\xi}_t :=  
  \begin{pmatrix} 
    A_t  \\ 
    X_t
  \end{pmatrix}
\end{equation}

\noindent satisfying the linear stochastic differential equation,

\begin{equation}
\label{linear_SDE}
	d \bm{\xi}_t  = \bm{Q}(t) \bm{\xi}_t  \, dt +  \bm{\sigma}(t) dW_t
\end{equation}

\noindent with the initial condition $\bm{\xi}_0^\mathsf{T} = \begin{pmatrix} 0 & 0 \end{pmatrix}$. $\bm{Q}(t)$ is a $2 \times 2$ matrix, $\bm{\sigma}(t)$ a 2 column matrix and $W_t$ is, as usual, a standard Wiener process. Furthermore, Chen and Georgiou consider that the two components $A_t$ and $X_t$ are linked by the relation,

\begin{equation}
\label{link_At_Xt}
	d A_t  = X_t  \, dt.
\end{equation}

\noindent which imposes a certain form on the matrix $\bm{Q}(t)$ and on the vector $\bm{\sigma}(t)$ as we will see shortly. The important point is the following: imposing a constraint on the final point $ A_T=A $ amounts to imposing the same constraint on the integral $ \int_0^T X_t \, dt = A $ which is precisely the constraint that we wish to impose on the paths of $X_t$. Thus, imposing a constraint $\bm{y}:= \begin{pmatrix} 
    A\\ 
    x_f
  \end{pmatrix} $ on the final state $\bm{\xi}_T$, 
\begin{equation}
\label{constraint_xi}
	\bm{\xi}_T =  \bm{y} \Leftrightarrow 
  \begin{pmatrix} 
    A_T \\ 
    X_T
  \end{pmatrix} 
                = 
  \begin{pmatrix} 
    A\\ 
    x_f
  \end{pmatrix} ,
\end{equation}

\noindent generates a bridge process $X_t$ having a fixed area under its path. Under those assumptions, the main result of Chen and Georgiou's article states that the stochastic process $\bm{\xi}_t$ conditioned on its final point $\bm{\xi}_T$ satisfies the stochastic differential equation~\cite{refChen}

\begin{equation}
\label{chen_SDE}
	d \bm{\xi}_t  = \left[ \left( \bm{Q}(t)  - \bm{\sigma}(t)\bm{\sigma}(t)^\mathsf{T}{\bm{P}}(t)^{-1} \right) \bm{\xi}_t +  \bm{\sigma}(t)\bm{\sigma}(t)^\mathsf{T}{\bm{P}}(t)^{-1} \bm{\Phi}(t,T) \bm{y} \right] \, dt +  \bm{\sigma}(t) dW_t,
\end{equation}

\noindent where ${\bm{P}}(t)$ is a $2 \times 2$ symmetrical matrix satisfying the differential Lyapunov equation

\begin{equation}
\label{eq_Lyapunov}
	\dot{{\bm{P}}}(t) = \bm{Q}(t) {\bm{P}}(t) + {\bm{P}}(t) \bm{Q}(t)^\mathsf{T} - \bm{\sigma}(t)\bm{\sigma}(t)^\mathsf{T}
\end{equation}
\noindent with the boundary condition ${\bm{P}}(T) = 0$ and where $\bm{\Phi}(t,\tau)$ is the state transition of Eq.~\eqref{linear_SDE} i.e. the matrix solution of the differential equation,
\begin{equation}
\label{eq_state_transition}
	\frac{d}{dt} \bm{\Phi}(t,\tau) = \bm{Q}(t) \bm{\Phi}(t,\tau) \qquad   \mathrm{with} \qquad \bm{\Phi}(\tau,\tau) = I_d \, ,
\end{equation}
\noindent where $I_d$ is the two-dimensional identity matrix. Compared to Eq.\eqref{linear_SDE}, observe that in Eq.\eqref{chen_SDE} only the drift term has changed (what we already knew from the theory of Doob~\cite{refDoob} and recovered in the formalism of Sottinen and Yazigi~\cite{refSottinen,refMazzolo}). Moreover, the new drift has two parts: a first term, proportional to $\bm{\xi}_t$, captures the behavior of the constraint along the paths and a second term, proportional to the constraints $\bm{y}$, sends the constraints on the desired endpoints $\bm{\xi}_T$ (this last term disappears if the constraints are fixed to zero). An identical behavior of the drift was obtained in the theory of Sottinen and Yazigi, however within their formalism the drift is given as a function of rather complicated stochastic integrals~\cite{refSottinen,refMazzolo}.\\

\noindent At first sight, the expression Eq.\eqref{chen_SDE} is not so simple since it requires the resolution of two equations, the Lyapunov equation and the matrix differential equation of the state transition. In fact, the two equations can be rather easily solved, giving its full strength to Chen and Georgiou's approach, as we will see now. However, before studying the general case of linear bridge process, we will first consider two important examples: the Brownian bridge and the Ornstein-Uhlenbeck bridge. In the following, for both examples and the general case, the vector of constraints will always be fixed at $\bm{y} = 
  \begin{pmatrix} 
    A\\ 
    x_f
  \end{pmatrix} $. For linear stochastic systems conditioned to a given final Gaussian probability distribution (instead of a point), see the recent series of papers by Chen, Georgiou and Pavon~\cite{refPavon1,refPavon2} and references therein. 

\subsection{Conditioned Brownian bridge}
In this session, we study the generalized Brownian bridge process conditioned on its area. By choosing, $\bm{Q}(t) = \begin{pmatrix} 
  0     & 1\\ 
  0     & 0
\end{pmatrix} $ and $\bm{\sigma}(t) = \begin{pmatrix} 
  0\\ 
  1
\end{pmatrix} $, Eq.~\eqref{linear_SDE} becomes 
\begin{equation}
  \begin{pmatrix} 
    dA_t  \\ 
    dX_t
  \end{pmatrix} =  \begin{pmatrix} 
  0     & 1\\ 
  0     & 0
\end{pmatrix}  
  \begin{pmatrix} 
    A_t  \\ 
    X_t
  \end{pmatrix} \, dt +  \begin{pmatrix} 
  0\\ 
  1
\end{pmatrix} dW_t ,
\end{equation}

\noindent or,
\begin{equation}
\label{integrated_brownian}
  \left\{
      \begin{aligned}
        & A_t = \int_0^t X_u \, du \\
        & d X_t = d W_t . \\
      \end{aligned}
    \right.
\end{equation}

\noindent Therefore, $X_t$ is a Brownian motion and $A_t$ is the so-called integrated Brownian motion. The two-dimensional process $\begin{pmatrix} A_t  \\ X_t \end{pmatrix} $ is often called the {\it{Kolmogorov diffusion}} since its study was initiated by Kolmogorov~\cite{refKolmogorov}. Let us add that Chen and Georgiou also studied this process with $\bm{y}^\mathsf{T}=\begin{pmatrix} 0 &0\end{pmatrix}$, by naming it, quite surprisingly, the Ornstein-Uhlenbeck bridge~\cite{refChen}. This point of vocabulary being specified, let us pursue the resolution of the problem. To this aim, let $a(t), b(t)$ and $c(t)$ be the coefficients of the symmetrical matrix ${\bm{P}}(t) = \begin{pmatrix} 
  a(t)     & b(t)\\ 
  b(t)     & c(t)
\end{pmatrix}$. With this set of parameters, the Lyapunov equation Eq.~\eqref{eq_Lyapunov} becomes
\begin{equation}
\label{eq_Lyapunov_integrated_brownian_bridge}
  \left\{ 
      \begin{aligned}
	 \dot{c}(t) & = -1   \\
	 \dot{b}(t) & = c(t)   \\
	 \dot{a}(t) & = 2 b(t) \, . \\ 
      \end{aligned}
    \right.
\end{equation}
\noindent Solving these differential equations with the boundaries condition $a(T) = b(T) = c(T) = 0$, we obtain, $c(t)=T-t$, $b(t)=-(T-t)^2/2$ and $a(t)= (T-t)^3/3$. Thus  ${\bm{P}}(t) = \begin{pmatrix} 
  \frac{(T-t)^3}{3}     & -\frac{(T-t)^2}{2} \\ 
  -\frac{(T-t)^2}{2}    & T-t
\end{pmatrix}$. Now, it remains to determine the state transition matrix $\bm{\Phi}(t,\tau)$ of Eq.~\eqref{chen_SDE}. However, for the integrated Brownian process, since the matrix $\bm{Q}(t)$ is time-invariant, $\bm{\Phi}(t,\tau)$ is given by the exponential of $(t-\tau)\bm{Q}$, that is

\begin{equation}
	\bm{\Phi}(t,\tau) = \exp[(t-\tau)\bm{Q}] = \exp \left[  (t-\tau) \begin{pmatrix} 
  0     & 1\\ 
  0     & 0
\end{pmatrix}  \right] = \begin{pmatrix} 
  1     & t - \tau\\ 
  0     & 1
\end{pmatrix} .
\end{equation}

\noindent Replacing these expressions in Eq.\eqref{chen_SDE} yields

\begin{equation}
\label{SDE_constrained_brownian_bridge_intermediate}
  \left\{
      \begin{aligned}
        & d A_t = X_t \, dt \\
        & d X_t = \left[-\frac{6}{(T-t)^2} (X_t - A) -\frac{2}{(T-t)} (2 X_t + x_f) \right] \, dt + d W_t  , \\
      \end{aligned}
    \right.
\end{equation}

\noindent and combining these two equations, we get

\begin{equation}
\label{SDE_constrained_brownian_bridge}
	d X_t = - \left[\frac{6}{(T-t)^2} \left(\int_0^t X_u \, du - A \right) +\frac{2}{(T-t)} (2 X_t + x_f) \right] \, dt + d W_t  . 
\end{equation}

\noindent Equation~\eqref{SDE_constrained_brownian_bridge} is the stochastic differential equation satisfied by a Brownian motion conditioned to end at $x_f$ and conditioned to have a fixed area $A$ under its curve. This equation can be found in~\cite{refMazzolo} while the zero area Brownian bridge, corresponding to the set of constraints $\bm{y}^\mathsf{T} = \begin{pmatrix} 
  0     & 0
\end{pmatrix} $, is derived in~\cite{refGorgens}. However, in both cases the stochastic differential equation of the constrained process was obtained by a substantially more technical approach. From Eq.\eqref{SDE_constrained_brownian_bridge}, the associate Langevin equation follows immediately,

\begin{equation}
\label{Langevin_constrained_brownian_bridge}
   \frac{dX_t}{dt} = - \frac{6}{(T-t)^2} \left(\int_0^t X_u \, du - A \right) - \frac{2}{(T-t)} (2 X_t + x_f)  + \eta(t) ,
\end{equation}

\noindent where $\eta(t)$ is a Gaussian white noise process~\cite{refMajumdarOrland}. Results presented in this section are not new but their obtaining shows how practical and efficient Chen and Georgiou's conditioning method is.

\subsection{Conditioned Ornstein-Uhlenbeck bridge}
In this session we study the generalized Ornstein-Uhlenbeck bridge process conditioned on its area.
By taking, $\bm{Q}(t) = \begin{pmatrix} 
  0     & 1\\ 
  0     & q
\end{pmatrix} $ with $q \in \mathrm{R} $ and $\bm{\sigma}(t) = \begin{pmatrix} 
  0\\ 
  1
\end{pmatrix} $, Eq.~\eqref{linear_SDE} becomes 
\begin{equation}
  \begin{pmatrix} 
    dA_t  \\ 
    dX_t
  \end{pmatrix} =  \begin{pmatrix} 
  0     & 1\\ 
  0     & q
\end{pmatrix}  
  \begin{pmatrix} 
    A_t  \\ 
    X_t
  \end{pmatrix} \, dt +  \begin{pmatrix} 
  0\\ 
  1
\end{pmatrix} dW_t ,
\end{equation}
\noindent or,
\begin{equation}
  \left\{
      \begin{aligned}
        & A_t = \int_0^t X_u \, du \\
        & d X_t = q X_t \, dt + d W_t , \\
      \end{aligned}
    \right.
\end{equation}

\noindent thus, $X_t$ is an Ornstein-Uhlenbeck process Eq.\eqref{OU-process-SDE}. With this set of parameters, the Lyapunov equation Eq.~\eqref{eq_Lyapunov} becomes

\begin{equation}
\label{eq_Lyapunov_Ornstein-Uhlenbeck_bridge}
  \left\{ 
      \begin{aligned}
	 \dot{c}(t) & = 2 \, q \, c(t) - 1   \\
	 \dot{b}(t) & = q \, b(t) + c(t)   \\
	 \dot{a}(t) & = 2 \, b(t) \, . \\ 
      \end{aligned}
    \right. 
\end{equation}

\noindent With the boundaries condition $a(T) = b(T) = c(T) = 0$, the solutions of this system of differential equations are

\begin{equation}
\label{eq_sol_Lyapunov_Ornstein-Uhlenbeck_bridge}
  \left\{ 
      \begin{aligned}
	 c(t) & =  \frac{1}{2 q} \left( 1 - e^{-2 q (T-t)} \right)  \\
	 b(t) & = -\frac{1}{2 q^2} \left( e^{-q (T-t)} - 1 \right)^2    \\
	 a(t) & = -\frac{1}{2 q^3} \left(3 - 4 e^{-q (T-t)} + e^{-2 q (T-t)} - 2 q (T-t) \right)  \, .\\ 
      \end{aligned}
    \right.
\end{equation}

\noindent Besides, since the matrix $\bm{Q}(t)$ is time-invariant, the state transition matrix $\bm{\Phi}(t,\tau)$ of Eq.~\eqref{linear_SDE} is given by the exponential of $(t-\tau)\bm{Q}$, that is,

\begin{equation}
\displaystyle 
	\bm{\Phi}(t,\tau) = \exp[(t-\tau)\bm{Q}] = \exp \left[  (t-\tau) \begin{pmatrix} 
  0     & 1\\ 
  0     & q
\end{pmatrix} \right] = \displaystyle  \begin{pmatrix} 
  1     & (-1+ e^{q (t - \tau)})/q\\ 
  0     & e^{q (t - \tau)}
\end{pmatrix} .
\end{equation}

\noindent Replacing these expressions in Eq.\eqref{chen_SDE} leads to

\begin{equation}
\label{SDE_constrained_Ornstein_Uhlenbeck_bridge_intermediate}
  \left\{
      \begin{aligned}
        d A_t & =  X_t \, dt \\
        d X_t & =  q \left[ \frac{-q \left( e^{q t}-e^{q T} \right)^2 (A_t - A) + \left( e^{2 q T}(1-q(T-t)) -e^{2 q t}(1+q(T-t))  \right)  X_t} {4 e^{q (t+T)}+e^{2 q T}(q (T-t)-2) -e^{2 q t} (q (T-t)+2)} \right.   \\
       & \left.  +  \frac{\left(e^{2 q t} - e^{2 q T} +2 q (T-t) e^{ q (T+t)} \right) x_f}{4 e^{q (t+T)}+e^{2 q T}(q (T-t)-2) -e^{2 q t} (q (T-t)+2)} \right] \, dt + d W_t . \\
      \end{aligned}
    \right.
\end{equation}

\noindent Combining these two equation and dividing by $dt$ gives the Langevin equation of the generalized Ornstein-Uhlenbeck bridge constrained to have a fixed area under its path

\begin{align}
\label{Langevin_constrained_Ornstein_Uhlenbeck_bridge}
        \frac{dX_t}{dt} & =  q \left[ \frac{-q \left( e^{q t}-e^{q T} \right)^2 (\int_0^t X_u \, du - A) + \left( e^{2 q T}(1-q(T-t)) -e^{2 q t}(1+q(T-t))  \right)  X_t} {4 e^{q (t+T)} +e^{2 q T}(q (T-t)-2) -e^{2 q t} (q (T-t)+2)}    \right] \nonumber \\
       & + \frac{q  \left(e^{2 q t} - e^{2 q T} +2 q (T-t) e^{ q (T+t)} \right) x_f}{4 e^{q (t+T)}+e^{2 q T}(q (T-t)-2) -e^{2 q t} (q (T-t)+2)}  + \eta(t) \, .  
\end{align}

\begin{figure}[!h]
\centering
\includegraphics[width=4.5in,height=3.2in]{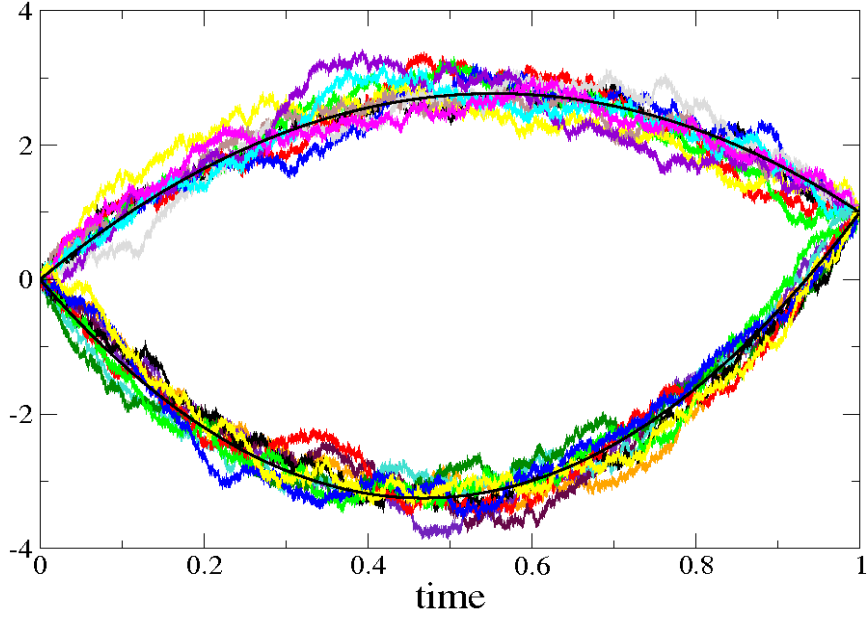}
\setlength{\abovecaptionskip}{15pt} 
\caption{Sample of Ornstein-Uhlenbeck paths conditioned to end at $x_f=1$ and conditioned to have a fixed area. Upper curves: the area is set at $2$. Lower curves: the area is set at $-2$. The time step used in the discretization is $dt=10^{-5}$. All trajectories generated with different noise histories are statistically independent. Thick black curves: average profile of the stochastic process as given by Eq.(\ref{constrained_OU-bridge-anticipative-mean}). Note that the asymmetry between the two curves appears when the final point $x_f$ is different from zero.}
\label{fig5}
\end{figure}

\noindent This equation simplifies considerably in the case of the conditioned Ornstein-Uhlenbeck bridge when $x_f=0$. Besides, although not obvious at first sight, the Langevin equation of the generalized conditioned Ornstein-Uhlenbeck bridge~Eq\eqref{Langevin_constrained_Ornstein_Uhlenbeck_bridge} is left invariant with respect to $q \to -q$ as for the Langevin equation of the Ornstein-Uhlenbeck bridge~Eq\eqref{non-anticipative-OU-bridge}. Figure~\ref{fig5} shows a set of 10 realizations of the process ending at $x_f = 1$ at time $T = 1$ for different fixed areas under the curve. Note that our approach is not limited to linear constraints; we can also choose global constraints on the path such as $\int_0^T X_t^2 dt$~\cite{refChetrite}. In this article, constraints on the final point and on the area have been studied, but in the technically easier case, either the first constraint or the second constraint is applied to the process, but not both together. \\

\vspace{1cm}

\subsection{Conditioned linear process}
\label{secLinearBridge}
We now turn our attention to the general case of the linear process. To this aim, we consider the matrices $\bm{Q}(t)$ and $\bm{\sigma}(t)$ of the form

\begin{equation}
 \bm{Q}(t) = \begin{pmatrix} 
  0     & 1\\ 
  0     & q(t)
\end{pmatrix} \qquad   \mathrm{and} \qquad  
\bm{\sigma}(t) = \begin{pmatrix} 
  0\\ 
  \sigma(t)
\end{pmatrix} ,
\end{equation}
\noindent where $q(t)$ and $\sigma(t)$ are continuous functions: $\mathbbm{R}^+ \to \mathbbm{R}$. With this choice of matrices, Eq.~\eqref{linear_SDE} becomes 
\begin{equation}
  \left\{
      \begin{aligned}
        & A_t = \int_0^t X_u \, du \\
        & d X_t = q(t) X_t \, dt + \sigma(t) d W_t , \\
      \end{aligned}
    \right.
\end{equation}
\noindent and with the additional constraints  $\begin{pmatrix}  A_T \\ X_T \end{pmatrix}  = \begin{pmatrix}  A \\ x_f \end{pmatrix} $  the process $X_t$ describes a linear bridge process with a fixed area. With this set of parameters the Lyapunov equation Eq.~\eqref{eq_Lyapunov} for the symmetrical matrix ${\bm{P}}(t) = \begin{pmatrix} 
  a(t)     & b(t)\\ 
  b(t)     & c(t)
\end{pmatrix}$
becomes,

\begin{equation}
\label{eq_Lyapunov_linear_bridge_bridge}  
  \left\{ 
      \begin{aligned}
	 \dot{c}(t) & = 2 \, q(t) \, c(t) - \sigma^2(t)   \\
	 \dot{b}(t) & = q(t) \, b(t) + c(t)   \\
	 \dot{a}(t) & = 2 \, b(t)  \, .\\ 
      \end{aligned}
    \right.
\end{equation}

\noindent With the boundaries condition ${\bm{P}}(T) = 0 \Leftrightarrow a(T) = b(T) = c(T) = 0$, the solution of this system of differential equations is given by,

\begin{equation}
\label{eq_sol_Lyapunov_linear_bridge_bridge}
  \left\{ 
\begin{aligned}
	c(t) & = e^{2\left(\bar{q}(t)- \bar{q}(T)\right)} \gamma(t,T) \\
     b(t) & = -e^{\bar{q}(t)- 2\bar{q}(T)} \int_t^T e^{\bar{q}(u)} \gamma(u,T) \, du \\
	a(t) & =  2 \int_t^T e^{\bar{q}(u)- 2\bar{q}(T)} \bar{\gamma}(u,T) ,\\ 
\end{aligned} 
  \right.
\end{equation}

\noindent where we have introduced the additional notations, partly borrowed from~\cite{refBarczy1},

\begin{align}
\label{eq_def_qbar_gamma_gammabar}
     \bar{q}(t)  & := \int_0^t q(u) \, du \nonumber \\
	\gamma(s,t) & := \int_s^t  e^{2\left[\bar{q}(t)- \bar{q}(u)\right]} \sigma^2(u) \, du  \\
     \bar{\gamma}(s,t) & := \int_s^t e^{\bar{q}(u)} \gamma(u,t) \, du .\nonumber 
\end{align}

\noindent As in the two preceding paragraphs, it remains to evaluate the state transition matrix $\bm{\Phi}(t,\tau)$ of Eq.~\eqref{chen_SDE}. Since the matrix $A(t)$ is no longer time-invariant, some extra work is needed to obtain the expression of the state transition matrix. This is done by standard techniques in appendix~\ref{appendix_2} where we find Eq.~\eqref{eq_state_transition_matrix},

\begin{equation}
\displaystyle 
	\bm{\Phi}(t,T)  = \displaystyle  
	\begin{pmatrix} 
		  \displaystyle 1     &  -e^{-\bar{q}(T)} \int_t^T e^{\bar{q}(u)} \, du \\ \\
		  \displaystyle 0     &\displaystyle   e^{- \bar{q}(T)+\bar{q}(t) }
	\end{pmatrix} .
\end{equation}

\noindent Furthermore, we have 
\begin{equation}
 \bm{\sigma}(t)\bm{\sigma}^\mathsf{T}(t) = \begin{pmatrix} 
  0     & 0\\ 
  0     & \sigma^2(t)
\end{pmatrix} \qquad   \mathrm{and} \qquad  
{\bm{P}}^{-1}(t)  = \frac{1}{b^2(t)-a(t)c(t)} \begin{pmatrix} 
  -c(t)     & b(t)\\ 
  b(t)      & -a(t)
\end{pmatrix} \, .
\end{equation}

\noindent Replacing these two expressions and that of $\bm{\Phi}(t,T)$ in Eq.\eqref{chen_SDE} leads to,

\begin{align}
\label{SDE_constraint_linear_bridge}
	d X_t & = \left[ \frac{- b(t) \sigma^2(t)}{b^2(t)-a(t)c(t)} \left( \int_0^t X_u \, du -A \right) +\left( q(t) +  \frac{a(t) \sigma^2(t)}{b^2(t)-a(t)c(t)} \right) X_t  \right] dt \nonumber \\
           & -x_f \frac{e^{-\bar{q}(T)}}{b^2(t)-a(t)c(t)} \left( a(t)e^{\bar{q}(t)} + b(t)\int_t^T e^{\bar{q}(u)} \, du \right) dt +  \sigma(t) dW_t ,
\end{align}

\noindent where the functions, $a(t),~b(t)$ and $c(t)$ are given by the Eqs.\eqref{eq_sol_Lyapunov_linear_bridge_bridge}. Equation~\eqref{SDE_constraint_linear_bridge} is the stochastic differential equation satisfied by a linear bridge conditioned to have a fixed area $A$ under its curve. Its associated Langevin equation follows immediately,

\begin{align}
\label{Langevin_constrained_linear_bridge_bridge} 
   \frac{dX_t}{dt} & =  \frac{- b(t) \sigma^2(t)}{b^2(t)-a(t)c(t)} \left( \int_0^t X_u \, du -A \right)  +\left( q(t) +  \frac{a(t) \sigma^2(t)}{b^2(t)-a(t)c(t)} \right) X_t \nonumber \\
                   & -x_f \frac{e^{-\bar{q}(T)}}{b^2(t)-a(t)c(t)} \left( a(t)e^{\bar{q}(t)} + b(t)\int_t^T e^{\bar{q}(u)} \, du \right) + \sigma(t) \eta(t) . 
\end{align}

\noindent For  $q(t)=q$ and $\sigma(t)= 1$, we recover the results of the previous paragraph Eq.~\eqref{Langevin_constrained_Ornstein_Uhlenbeck_bridge}, besides the Langevin equation simplifies significantly in the bridge case when $x_f = 0$. In addition, in Appendix~\ref{appendix_1}, when a single constraint is imposed on the process, we establish the link between the formalism used in this paper and other approaches.

\section{Conclusion}
\label{sec4}

Conditioned Ornstein-Uhlenbeck processes, and more generally linear processes, are usually achieved by Doob's {\it{h}}-transform and its refinements. However, global constraints cannot be easily handled within Doob formalism. To overcome this difficulty, we presented in this article two different approaches, the first one based on some recent results obtained by Sottinen and Yazigi~\cite{refSottinen} for conditioned Gaussian processes and the second one, by Chen and Georgiou~\cite{refChen} has its roots in the theory of stochastic control. In particular, from Sottinen and Yazigi's technique, we derived an anticipative representation of the generalized Ornstein-Uhlenbeck bridge conditioned to have a fixed area under its curve. Constraints on the stochastic process appear clearly in this anticipative representation. Moreover, the mean behavior of the process, as well as its variance, come easily. However, the anticipative representation requires knowledge of the future and consequently is not entirely satisfactory. On the contrary, the non-anticipative representation, obtained from Chen and Georgiou's approach, is adapted to the filtration but the constraints are somewhat hidden in the drift term (which can be rather intricate). Nevertheless, this representation has one important advantage: Trajectories are efficiently generated by a Langevin equation and are all independent. Besides, the Chen and Georgiou's method of conditioning is quite simple and possible extensions to linear stochastic systems conditioned to a given final probability distribution (instead of a point) are currently an active field of research.

\section{Acknowledgements}
The author wish to thank Dr. Fausto Malvagi for reading the manuscript and his comments.

\appendix
\section{Linear bridges: link with the formulation of Barczy and Kern}
\label{appendix_1}
In a recent article, Barczy and Kern~\cite{refBarczy1} have studied a linear process $Z_t$ given by the stochastic differential equation,

\begin{equation}
  \left\{
      \begin{aligned}
        & d Z_t = q(t) Z_t \, dt + \sigma(t) d W_t , \\
        & Z_0 = 0 \\
      \end{aligned}
    \right.
\end{equation} 

\noindent conditioned to end at $x_f$ at time $T$ (we call $U_t$ this constrained process). The process $U_t$ is the linear bridge encountered in section Sec~\ref{secLinearBridge}. Using transition densities of $Z_t$, they have obtained the stochastic differential equation of the linear bridge,

\begin{equation}
\label{SDE_kern_barczy}
        d U_t = \left[ q(t) -\frac{e^{2\left(\bar{q}(T)- \bar{q}(t)\right)} }{\gamma(t,T)} \sigma^2(t) \right] U_t \, dt + x_f \frac{e^{\bar{q}(T)- \bar{q}(t)} }{\gamma(t,T)} \sigma^2(t) \, dt + \sigma(t) d W_t .
\end{equation}

\noindent In this appendix, we will recover these results in the light of Chen and Georgiou's approach. The linear bridge case is simpler than those studied so far since the process has only one constraint (on the final point). Consequently, the process is scalar as well as the associated stochastic differential equation Eq.~\eqref{linear_SDE} which is written,
\begin{equation}
\label{linear_scalar_SDE}
	d \xi_t  = Q(t) \xi_t  \, dt +  \sigma(t) dW_t ,
\end{equation}

\noindent where $Q(t)= q(t)$ and $ \sigma(t)$ are two scalar functions of time. The stochastic differential equation satisfies by the linear bridge, i.e. the process $\xi_t$ with the constraint $\xi_T = 0$, is given by Eq.~\eqref{chen_SDE},

\begin{equation}
\label{chen_SDE_simple_linear_bridge}
	d \xi^c_t  =  \left[ \left( Q(t)  - \frac{\sigma^2(t)}{P(t)} \right) \xi^c_t + \frac{\sigma^2(t)}{P(t)} \Phi(t,T) x_f \right] \, dt +  \sigma(t) dW_t \, .
\end{equation}

\noindent The scalar function $P(t)$ satisfies the Lyapunov equation Eq.~\eqref{eq_Lyapunov},
\begin{equation}
\label{eq_Lyapunov_simple_linear_bridge}
	\dot{P}(t) = 2q(t) P(t)  - \sigma^2(t),
\end{equation}
\noindent with the boundary condition $P(T) = 0$. The solution of this differential equation is given by
\begin{equation}
\label{eq_P_simple_linear_bridge}
	P(t) = e^{2 \bar{q}(t)} \int_t^T e^{-2 \bar{q}(u)} \sigma^2(u) \, du = \gamma(t,T) e^{2 \left( \bar{q}(t)- \bar{q}(T) \right)} .
\end{equation}

\noindent The functions $\bar{q}(t)$ and $\gamma(s,t)$ were defined in~Eq.\eqref{eq_def_qbar_gamma_gammabar}. The state transition $\Phi(t,T)$ satisfies the differential equation Eq.~\eqref{eq_state_transition}, 
\begin{equation}
\label{eq_state_transition_linear}
	\frac{d}{dt} \Phi(t,\tau) = q(t) \Phi(t,\tau) \qquad   \mathrm{with} \qquad \Phi(\tau,\tau) = 1\, ,
\end{equation}
\noindent whose solution is $\Phi(t,\tau) = e^{\int_{\tau}^t q(u) \, du } = e^{\bar{q}(t)- \bar{q}(\tau)}$. Replacing this expression and that of $P(t)$ in Eq.~\eqref{chen_SDE_simple_linear_bridge} leads immediately to the result of Barczy and Kern Eq.~\eqref{SDE_kern_barczy}. This example shows that the method presented in this article is a real shortcut to obtain the stochastic differential equations of constrained processes. Note that in the case of a linear bridge, the solution of the stochastic differential equation Eq.~\eqref{SDE_kern_barczy} is known~\cite{refBarczy1},

\begin{equation}
\label{sol_SDE_simple_linear_bridge}
  \left\{
      \begin{aligned}
        U_t & = \frac{\gamma(0,t)}{\gamma(0,T)}  e^{\bar{q}(T)-\bar{q}(t)} x_f   + \int_0^t \frac{\gamma(t,T)}{\gamma(s,T)}  e^{\bar{q}(t)-\bar{q}(s)} \sigma(s) \, dW_s \qquad  0 \leq t < T \\
        U_T & = x_f  .  \\
      \end{aligned}
    \right.
\end{equation}

\noindent When $q(t)=q \neq 0$ and $\sigma(t) =\sigma \neq 0$, corresponding to the Ornstein-Uhlenbeck bridge, the solution takes the form,

\begin{equation}
\label{sol_SDE_simple_ornstein_uhlenbeck_bridge}
  \left\{
      \begin{aligned}
        U_t & = \frac{\sinh\left( q t\right)}{\sinh\left(q T\right)} x_f   + \  \sigma   \int_0^t \frac{\sinh\left(q(T-t)\right)}{\sinh\left(q(T-s)\right)}  \, dW_s \qquad  0 \leq t < T \\
        U_T & = x_f  , \\
      \end{aligned}
    \right.
\end{equation}
\noindent an expression we encountered in section~\ref{sec2}.

\section{state transition matrix}
\label{appendix_2}
In this appendix, following a standard technique described in~\cite{ref_book_Aplevich}, we give a closed form of the state transition matrix associated with the equation, 

\begin{equation}
   \frac{d}{dt} \bm{X}(t) = \bm{Q}(t) \bm{X}(t)  \qquad \mathrm{with} \qquad
 \bm{Q}(t) = \begin{pmatrix} 
  0     & 1\\ 
  0     & q(t)
\end{pmatrix} \, ,
\end{equation}

\noindent $\bm{X}(t)$ being a two-component vector $\begin{pmatrix} X_1(t) \,  X_2(t) \end{pmatrix}$. The system is equivalent to the equations,

\begin{equation}
\label{eq_x1_and_x2}
  \left\{ 
      \begin{aligned}
	 \dot{X_1}(t) & = X_2(t)       \\
	 \dot{X_2}(t) & = q(t) X_2(t)   
      \end{aligned}
    \right. \qquad \Rightarrow 
   \left\{ 
      \begin{aligned}
	 \dot{X_1}(t) & = C_1 \int_0^t e^{\bar{q}(u)} \, du + C_2    \\
	 \dot{X_2}(t) & = C_1 e^{\int_0^t q(u) \, du} = C_1 e^{\bar{q}(t)}
      \end{aligned}
    \right. \, ,
\end{equation}

\noindent where $C_1$ and $C_2$ are two constants. Choosing the initial conditions $\bm{X}(0) = \begin{pmatrix} 
  1 \\ 
  0
\end{pmatrix} $ 
gives the solution $\bm{\Psi}_1(t) = \begin{pmatrix} 
  1 \\ 
  0
\end{pmatrix} $ 
and choosing the initial conditions  $\bm{X}(0) = \begin{pmatrix} 
  0 \\ 
  1
\end{pmatrix} $ 
gives the solution $\bm{\Psi_2}(t) = \begin{pmatrix} 
  \int_0^t e^{\bar{q}(u)} \, du  \\ 
  e^{\bar{q}(t)}
\end{pmatrix} $ . 
\noindent Thus, a fundamental matrix is $\bm{\Psi}(t) = \begin{pmatrix} \bm{\Psi}_1(t)  \, \bm{\Psi}_2(t) 
\end{pmatrix} = 
\begin{pmatrix} 
  1     &  \int_0^t e^{\bar{q}(u)} \, du \\ 
  0     & e^{\bar{q}(t)}
\end{pmatrix}
$ and therefore the state transition matrix is, 

\begin{equation}
\label{eq_state_transition_matrix}
\displaystyle 
	\bm{\Phi}(t,T)  = \bm{\Psi}(t) \bm{\Psi}^{-1}(T) =  \begin{pmatrix} 
  1     &  \int_0^t e^{\bar{q}(u)} \, du \\ 
  0     & e^{\bar{q}(t)}
\end{pmatrix} 
\displaystyle 
\begin{pmatrix} 
  1     &  \int_0^T e^{\bar{q}(u)} \, du \\ 
  0     & e^{\bar{q}(T)}
\end{pmatrix} ^{-1} =
\displaystyle  
	\begin{pmatrix} 
		  \displaystyle 1     &  -e^{-\bar{q}(T)} \int_t^T e^{\bar{q}(u)} \, du \\ \\
		  \displaystyle 0     &\displaystyle   e^{- \bar{q}(T)+\bar{q}(t) }
	\end{pmatrix} .
\end{equation}


\end{document}